\documentclass[a4paper,12pt]{article}
\pdfoutput=1
\usepackage{amssymb}
\usepackage{amsmath}
\usepackage{epsfig}
\usepackage{graphicx}
\usepackage{color,ulem}
\usepackage{hyperref}

\makeatletter

\@addtoreset{equation}{section}
\makeatother
\def\be{\begin{equation}}
\def\ee{\end{equation}}
\def\bea{\begin{eqnarray}}
\def\eea{\end{eqnarray}}
\def\eq{\begin{eqnarray}}
\def\eqx{\end{eqnarray}}

\def\({\left(}
\def\){\right)}
\def\<{\left<}
\def\>{\right>}

\def\be{\begin{equation}}
\def\ee{\end{equation}}
\def\ben{\begin{eqnarray}}
\def\een{\end{eqnarray}}
\def\({\left(}
\def\){\right)}
\def\<{\left<}
\def\>{\right>}
\def\!{\right|}
\def\|{\left|}

\def\[{\left[}
\def\]{\right]}

\def\+{\bar}

\def\M{{\cal{M}}}

\def\bx{{\bf{x}}}

\newcommand{\lambdabar}{{\mkern0.75mu\mathchar '26\mkern -9.75mu\lambda}}

\usepackage{hyperref}
\hypersetup{colorlinks,%
  citecolor=blue,%
  linkcolor=blue,%
  pdftex}

\begin{document}

\begin{titlepage}
\vskip1cm
\begin{flushright}
\end{flushright}
\vskip0.25cm
\centerline{
\bf \large
Effective 
Cross Section of Fuzzy Dark Matter Halos
}
\vskip1cm \centerline{ 
 Dongsu Bak,$^{\, \tt a,c}$  Jae-Weon Lee,$^{\, \tt b}$ Sangnam Park$^{\, \tt c}$} 
\vspace{1cm}
\centerline{\sl  a) Physics Department,
University of Seoul, Seoul 02504 \rm KOREA}
 \vskip0.3cm
 \centerline{\sl b) Department of Electrical and Electronic Engineering, Jungwon University}
 \centerline{85 Munmuro, Goesan, Chungbuk 28024 \rm KOREA}
   \vskip0.3cm
 \centerline{\sl c) Natural Science Research Institute,
University of Seoul, 
Seoul 02504 \rm KOREA}
 \vskip0.4cm
 \centerline{
\tt{(dsbak@uos.ac.kr,\,scikid@jwu.ac.kr,\,u98parksn@gmail.com})}
  \vspace{2cm}
\centerline{ABSTRACT} \vspace{0.75cm}
{
\noindent
 We   numerically study  the movement of two colliding fuzzy dark matter solitons without explicit self-interaction
 and find the effective cross section of dissipative  change in velocity.
 The  cross section turns out to be inversely proportional to the velocity cubed,
and we present its
analytic interpretation.
Using the result we roughly estimate  spatial offsets during head-on  collisions of two fuzzy dark matter halos, which can be related to
 the spatial offsets between stars and dark matter in collisions of some galaxy clusters.
  We also show that the gravitational cooling plays an important role during the collisions.
}
\end{titlepage}

\section{Introduction}\label{sec1}

Numerical simulations with the cold dark matter (CDM) models  are  successful in explaining the
cosmic large scale structures; however,
the CDM models encounter small scale issues  in explaining galactic structures~\cite{Navarro:1995iw,deblok-2002,crisis}.
For example, numerical studies with the CDM models predict a cusped galactic halo central  density, which seems to be inconsistent with  observational data showing flat core-like densities in small galaxies.

Recently, as an alternative to the CDM models, there is a growing interest in the fuzzy DM model~\cite{Fuzzy,Hui:2016ltb,Lee:2017qve}
in which the DM is a scalar particle with the ultra-light mass $m\simeq  10^{-22} {\rm \, eV}$ in the state of
Bose-Einstein condensate. 
This DM is also often called scalar field DM, wave DM, or ultra-light axion.
In this model the long  Compton wavelength $\lambda_c=2\pi \hbar/mc\simeq 0.4 {\rm \,pc} $
and the wave nature of the condensed particles help us to resolve the small-scale issues.

However, the effects of visible matter on the structure formation  in larger galaxies are complicated and non-negligible. This fact renders interpretation of  observational data to discriminate DM models difficult.
On the other hand, since DM is more dominant at the center of galaxy clusters  than
at the center of large galaxies, the clusters can provide a good testbed to study the nature of DM.

For example, one of the mysteries of the clusters lies in the spatial offsets between the luminosity peaks (stars) and the peaks of the DM density inferred
from gravitational weak lensing at central regions. 
In the simple collision-less CDM models, DM particles have no self-interaction besides gravitational interaction, and
stars and DM should
move together during merging. However, observations indicate apparent offsets in some clusters, which can give us
 constraints on  self-interaction parameters of DM.
 The self-interaction scattering cross section per unit mass  $\sigma/m$ should be about $1 {\rm \, cm^2/g}$
 to solve the small scale issues, while the
constraints arising from the small offsets during cluster collisions imply  $\sigma/m <{\cal O}(0.1){\rm \, cm^2/g}$~\cite{Harvey:2015hha}.
There are some galaxy clusters like Abel 3827 that need a larger cross section   $\sigma/m > 1{\rm \,cm^2/g}$ to explain their offsets~\cite{Tulin:2017ara}.
This apparent contradictive behavior can be explained with a velocity-dependent cross section in the self-interacting DM model or
with
the soliton-like nature in the fuzzy DM
model~\cite{lee2008bec,Paredes:2015wga,Guzman:2016peo}.
To study the  spatial offsets in the fuzzy DM model we need to study the collision of DM halos, which can be
 modeled as fuzzy DM solitons
 satisfying the Schr\"odinger-Poisson equation.

The Schr\"odinger-Poisson equation has
 the gravitational cooling \cite{Guzman:2006yc, Bak:2018pda}, which is a mechanism for relaxation
by ejecting part of the fuzzy DM and carrying out excessive kinetic energy.
During  collisions two fuzzy DM halos have interfering field profiles, which inevitably contain high momentum modes.
These modes can escape the gravitational potential of the halos, which leads to reduction of the velocities of the  halos.

 The aim of this paper is to find the effective cross section of two colliding fuzzy DM solitons without explicit self-interaction terms
and to roughly estimate  spatial offsets during head-on  collisions of two fuzzy DM halos.
  We will show that the gravitational cooling plays an important role during the collisions.

In  Sec.~{\ref{sec2}} we present  results of our numerical study for the collisions
and theoretical interpretation of the results.
In  Sec.~{\ref{sec3}} we apply the results to  estimate the  spatial offsets during a collision of two model galaxy clusters.
The last section is devoted to concluding remarks.

\section{Collisions of solitonic cores}\label{sec2}
Galaxy clusters are roughly consisting of dark matter halos ($\sim$ $85\%$), intra-cluster gases ($\sim 10\%$), and the remaining fraction of compact objects (mainly stars) made out of ordinary matters \cite{Bahcall:1995tf}.
In our model, the dark matter halos mainly  consist of solitonic cores where each solitonic core may be considered as a ground state of Schr\"odinger-Poisson system \cite{Diosi:2014ura, Moroz:1998dh}. These solitonic cores have typically masses of about $10^8 M_{\odot}$ with a size of about one $\text{kpc}$ scale. Indeed in the simulation of cosmological
structure formation with our fuzzy DM \cite{Schive:2014dra}, granular structures have been  numerically found where each granule may be considered as a solitonic core of the  Schr\"odinger-Poisson system. In this section we shall consider collisions of  two such solitons
in the Schr\"odinger-Poisson system
\begin{align}    \label{}
i\hbar \partial_t \psi(\bx,t ) & = -\frac{\hbar^2}{2m}\nabla^2 \psi(\bx,t )  +m V(\bx,t ) \psi(\bx,t )  
\\
 \nabla^2 V(\bx,t )&=4\pi GM_{\rm tot} \, |\psi|^2(\bx,t )
\end{align}
where $M_{\rm tot}$ is the total mass of the system and  the wave function is normalized by
$\int d^3 \bx\,  |\psi|^2 
=1$.
There is so far no analytic understanding of this collision problem. It is rather complicated even with current technologies of numerical simulation due to its four dimensional nature which in general requires a large amount of CPU time. With a rather limited goal of accuracy, we shall carry out numerical analysis of  the collision problem. In order to map into the code space, we introduce dimensionless variables by the rescaling
\begin{align}    \label{}
t & \rightarrow   \tau_c 
\, t   = \frac{\hbar^3}{m^3} \frac{1}{\left( G\M\right)^2} t \\
\bx &  \rightarrow   \ell_c 
\, \bx   = \frac{\hbar^2}{m^2} \frac{1}{G\M} \, \bx\\
\psi & \rightarrow   \alpha_\psi \, \psi   = \frac{m^3}{\hbar^3} \left( G\M\right)^{\frac{3}{2}} \left(\frac{M_{\rm tot}}{\M}\right)^{\frac{1}{2}} \psi \\
V& \rightarrow   \alpha_V \, V   = \frac{m^2}{\hbar^2} \left({4\pi G\M}\right)^{2}  V
\end{align}
leading to a form 
suited for the numerical analysis
\begin{align}    \label{SchEq}
i \, \partial_t  \psi(\bx,t ) \, & = -\frac{1}{2}\nabla^2 \psi(\bx,t )  + V(\bx,t ) \psi(\bx,t )  
\\
 \nabla^2 V(\bx,t )&=\ 4 \pi \, |\psi|^2(\bx,t ) \label{codeEq}
\end{align}
Note that the normalization of the wave function in the code space becomes $M_{\rm tot}= \M \int d^3 \bx\,  |\psi|^2 $ where ${\cal M}$ is our code-unit mass scale that in turn determines the unit-time  ($\tau_c$) and the unit-length scale ($\ell_c$)  in our code space\footnote{Thus $\int d^3 \bx\,  |\psi|^2$ in the code space is no longer fixed to be unity in general.}.  We shall take     this code-unit  mass scale as $\M = 4\pi \times 10^7 M_\odot$ and
$m=10^{-22} \text{eV}/c^2$.  This then fixes the time and length scales as
 $\tau_c  \sim 2.3584 \times 10^7 \, \text{yr}$ and
 $\ell_c  \sim 0.68000\,  \text{kpc}$ respectively. Hence the unit velocity in the code space corresponds to $v_c = \ell_c/\tau_c \sim    28.194 \, \text{km/s}$ in the real space.

We shall use the python  package, PyUltraLight, developed in \cite{Edwards:2018ccc}, whose source code is publicly available. In this package,  the system is placed
in a box of size $L$ with the spatially  periodic boundary condition. 
We now consider a head-on collision of two solitons with an equal  soliton mass $M$.\footnote{ Hence $M_{\rm tot}=2M$ in our setup.} Once they are sufficiently well separated, each core may be approximated by a spherically symmetric, solitonic ground-state solution of the
Schr\"odinger-Poisson system. In particular, its half-mass radius may be
estimated as
$r_{\frac{1}{2}}(M)= f_0\ell_c {\M}/{M} $
with $f_0 \sim 3.9251$ \cite{Hui:2016ltb}. In the following, we shall refer this scale as a size of  soliton. As an initial configuration, we shall put two such solitons at $(x_L=-\frac{1}{4}L,0,0)$
and $(x_R=+\frac{1}{4}L,0,0)$ with  velocities $v_L= -v_R=\frac{1}{2}v_i$ in the $x$ direction respectively where $v_i$ denotes their initial relative velocity\footnote{For each moving solition, one may introduce an additional velocity-dependent phase factor \cite{Edwards:2018ccc} and we shall put the relative phase factor  $\delta=0$ for the simplicity of our analysis (See \cite{Edwards:2018ccc} for the details).}. The number of grid point $N_g$ in each direction is taken to be $400$. Hence our spatial resolution has been given by $\Delta x= {L}/{N_g}$.  The default time step of the program
is given by $\Delta t_{\rm d} = \frac{\Delta x^2}{\pi}$. In our problem, we use instead
the time step $\Delta t = \frac{{\rm Duration}} {N_t}$ with the number of
time step $N_t = 4000$ where the duration is a rough estimation of the time required for the left soliton to reach $x_L = \frac{1}{4}L$ from its initial position. It turns out
that for all of our simulations, $\Delta t < \frac{2}{3}\Delta t_{\rm d}$.
This choice seems better suited for our simulation where
 velocity dependence plays an important role.
In this head-on collision, we find that the two solitons pass through each other unless they merge\footnote{Labeling each soliton could be ambiguous since the collision is solely described by field profiles of our Schr\"odinger-Poisson system. However this ambiguity may be lifted by considering slightly different masses, {\it e.g.} $M_L > M_R$, and tracking for instance the larger/smaller density peak as the soliton L/R.}.
We then measure the final velocity after the collision at $x_L = \frac{3}{16} L$.  We then compute initial/finial  relative velocities at infinity using the energy conservation
\bea
\frac{1}{4} M v^2_{\infty}=\frac{1}{2}M v_{L\infty}^2 +\frac{1}{2}M v_{R\infty}^2     = \frac{1}{2}M v_L^2 + \frac{1}{2}M v_R^2 - \frac{M^2}{|x_R\negthinspace -\negthinspace x_L|}
\eea
This way the change of the relative velocity    $\Delta v \equiv  v^f_{\infty}-v^i_{\infty}$ can be computed for each simulation. For simplicity of our presentation, the initial relative velocity  $v^i_{\infty}$ shall be denoted simply by $v$ in the following.
We carry out simulation
with $M=23,24,25,26,27,28$ for $L=20$.  For the choice of $L=10$, we have
$M=40,42,44,46,48,50$. The initial relative velocities are chosen as
$v_i =36,38,40,42,44,46,48,50$. Thus one has 96 data points in total.

\begin{figure}[thb]
\begin{center}
\includegraphics[width=0.5\textwidth]{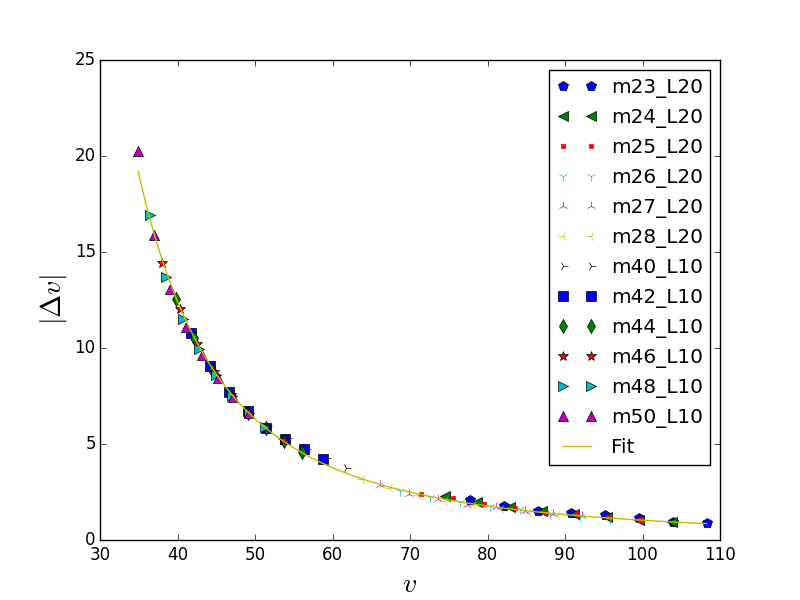}\includegraphics[width=0.5\textwidth]{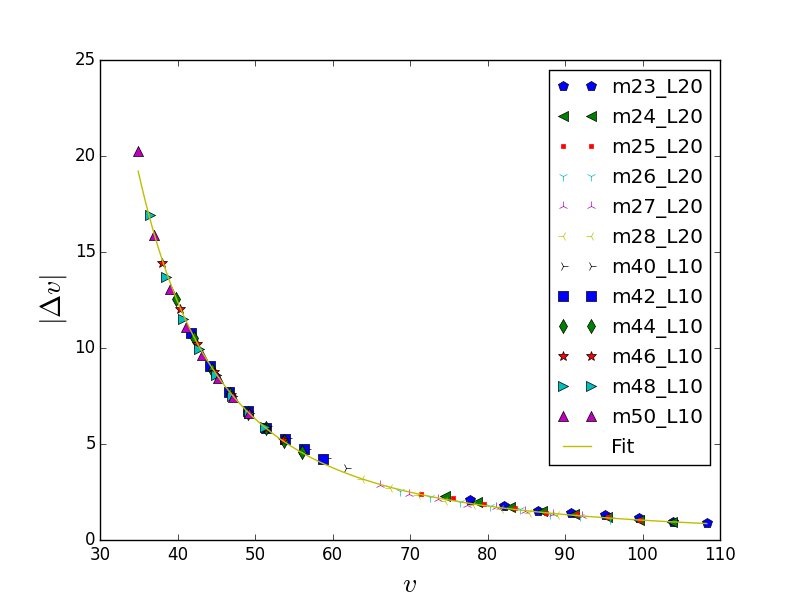}
\caption{ On the left, we depict the $3$-parameter fit of our $96$ data points where fitting function is given by $|\Delta v|= \alpha v (M_0/v)^{\lambda} (1+ \beta (M_0/v)^2)$ with $M_0=50$. On the right, we fix $\lambda=3$ and fit the same data 
with the $2$-parameter fit function
 $|\Delta v|= \alpha v (M_0/v)^{3} (1+ \beta (M_0/v)^2)$.  }
\label{dv_v_plot}
\end{center}
\end{figure}

Now note that our problem involves only two independent length scales: one is the half mass radius of each soliton given by $r_{1/{2}}$ and the other is the de-Brogile wavelength (divided by $2\pi$) of a mass $m$ particle moving with velocity $v$ that is given by $\lambdabar (v)=\frac{\hbar}{mv}$.  Since $|\Delta v|/v$ is dimensionless, it should be a function of the combination $\frac{\hbar}{mv \, r_{{1}/{2}}}$ only. Therefore, one has a general form
\be
|\Delta v | = v \, G_c(M/v)
\ee
where $G_c$ is the function of the combination $M/v$ only.
Note that $G_c(z)/z$ should be even function of $z$ which follows from the symmetry of the problem under the parity transformation $x \rightarrow -x$.
With this scaling form, one can send   the soliton-mass parameter $M$ to a fixed value $M_0 = 50$ with the scaling
$v  \rightarrow   v'= v M_0/M$  
and
 $\Delta v  \rightarrow  \Delta v'= \Delta v M_0/M$. 
This way,  we generate a new data point $(v', \Delta v')$ with $M_0 =50$ for any given data point. Hence finally one has $96$ data points with $M_0=50$ which are depicted in Figure \ref{dv_v_plot}. Let us now fit the function $G_c(z)$ with $z=M_0/v$.
With a 3-parameter fit of the form
\be
G_c(z) = \alpha z^{\lambda} \left(1+ \beta{z^2} \right)
\ee
whose choice will be further justified below, we find that the leading power index $\lambda=
3.0106\pm 0.3240$, $\alpha=0.07152\pm 0.11321$ and $\beta=0.8405\pm 0.5728$. The result is depicted on the left side of Figure \ref{dv_v_plot}. In particular, we find that the power index $\lambda$ approaches $3$ as we add more and more data points to the fitting. We then set $\lambda = 3$ and try the $2$-parameter fit with the fitting function
\be
G_c(z) =\alpha z^{3} \left(1+ \beta z^2 \right)
\ee
This leads to $\alpha=0.06791\pm 0.00114$ and $\beta=0.8592\pm 0.0265$ and the corresponding fit function is depicted on the right side of Figure  \ref{dv_v_plot}.

Let us now argue for the leading cubic power as follows. This leading power will become precise
in the large velocity limit where any possible  dissipation including the gravitational cooling effect \cite{Guzman:2006yc, Bak:2018pda}
 becomes negligibly small. In this part of discussions, we shall focus on the large velocity limit of
the head-on collision of two solitons with the same mass $M$. Assuming there were no dissipation,  the masses of two solitons would be preserved after the collision once they are separated by an enough distance, and thus
 $\Delta v$ would vanish  due to the total energy conservation.
This can also be understood from the symmetry of the problem as follows. In the $x$-directional head-on collision, we only need to consider the momentum transfer $\Delta p_L$ of the left body by the right in the $x$-directions. Then this momentum transfer would  be given by
\bea
\Delta p_L = \int^{\infty}_{-\infty} dx_{r} \frac{dt}{dx_{r}} F(x_{r})
\eea
where $x_r(t)=x_L(t)- x_R(t)$ and $F(x_r(t))$ is the $x$ directional effective force on the left body by the right. We now note that the velocity $dx_r(t)/dt$ and  $F(x_r(t))$ would  respectively be even and odd under the transformation $t\rightarrow -t$ with the choice $x_r(0)=0$.
Therefore $\Delta p_L =M \Delta v = 0$ with the symmetry of the problem. This symmetry will be broken mainly by the gravitational cooling effect.  By any excitation from a stationary state  of our Schr\"odinger-Poisson system, the perturbation will in general produce an out-going flux of probability proportional to $1/\tau$ where $\tau=\tau_c \frac{\M^2}{M^2}$ is the time scale of any relaxation processes in our soliton system, which is nothing but the gravitational cooling effect. Since the main part of perturbation occurs when two solitons are overlapping significantly, the duration is given by $\Delta \tau_{\rm cross}=\frac{2 r_{1/2}}{v}$.   Hence the dissipated fraction by the gravitational cooling effect will be proportional to $\frac{\Delta \tau_{\rm cross}}{\tau}$, whose main effect is to reduce the velocity $dx_r/dt$ of solitons in the collisional process.  Thus
the change in the momentum can be estimated as follows\footnote{We would like to thank the referee for  clarification of the argument below.}:
First note that, while two solitons are overlapping with each other, the velocity may be estimated by
\be
 \frac{dx_{r}}{dt} \sim \Bigl(\frac{dx_{r}}{dt} \Bigr)_0\Bigl(1-\delta' \, \frac{\Delta \tau_{\rm cross}}{\tau} \frac{x_r}{r_{1/2}}  +{\cal O} \bigl(\frac{\Delta \tau^2_{\rm cross}}{\tau^2}\bigr)\Bigr)
\ee
where $\Bigl(\frac{dx_{r}}{dt} \Bigr)_0$ denotes the velocity before inclusion of any dissipative effect and  $\delta'$ is an order-one numerical constant. We here omit any parity-even
contribution in the leading correction 
since it  does not contribute to an  evaluation of the
following integral. The change in momentum then becomes
\be
\Delta p_L \propto \int dx_{r} \Bigl( \frac{dt}{dx_{r}}\Bigr)_0 \Bigl(1+\delta' \, \frac{\Delta \tau_{\rm cross}}{\tau} \frac{x_r}{r_{1/2}} \Bigr) F(x_{r}) \propto -\frac{\Delta \tau_{\rm cross}}{\tau} \frac{GM^2}{r^2_{1/2}}
\Delta \tau_{\rm cross}
\ee
where the order-one contribution of the integral vanishes as was mentioned previously.
From this, one finds
\be
\Delta v /v= -\alpha'   \left(\frac{\hbar}{m v r_{1/2}}\right)^3
\label{leadingratio}
\ee
where we used the relation $\Delta v /v=\Delta p_L/p_L$ and $\alpha'$ is an order-one numerical constant. Note that, in this argument, the presence of the gravitational cooling effect plays a rather crucial
role.

We now translate our numerical result in terms of the 
variable
$q=\frac{\hbar}{mv r_{1/2}}$. Then the $2$-parameter fit function becomes
\be
|\Delta v| /v=   G(q) =A q^{3} \left(1+ B q^{2} \right)
\label{deltav}
\ee
where $A=4.106\pm 0.069$ and $B=13.24\pm 0.41$\footnote{The parameters $A, B$ are related to $\alpha,\beta$ by
$A=f_0^3 \, \alpha$ and $B=f_0^2\, \beta$.}. This will be used  for the estimation of the offset  in the next section.

Now let us discuss validity regime of the above results. When $q\gg 1$, the de Brogile wavelength becomes much larger than the soliton size $r_{1/2}$, which corresponds to so-called
``classical" regime \cite{Hui:2016ltb,Lancaster:2019mde} of  scattering processes.  In this classical regime, our solitons behave like classical particles when they are well separated. In our head-on-collision problem however, two solitons in the classical regime merge due to the dissipative effect  as will be explained further below.
The other limit $q\ll 1$ is called  ``quantum" regime \cite{Hui:2016ltb,Lancaster:2019mde}, where the wave nature of profiles becomes more prominent and, upon colliding,  solitons pass through each other like
usual colliding wave-packets.

In our simulation in the above, $q$ variable is ranged over
$[0.117, 0.353]$ and in the application in Section \ref{sec3}, $q$ will be assumed to be smaller than
 ${q}_{\rm cutoff}=0.282$ where ${q}_{\rm cutoff}$ is for $M\simeq 0.5f_0{\cal M}$ and $v=50 \, {\rm km/s}$. Thus, in this note, we shall be mostly working in the quantum regime. In fact, once $|\Delta v|/v$ reaches $1$, the two solitons merge with each other and thus the region of $|\Delta v|/v >1$ is not allowed. Hence $G(q)$ should be replaced by $G(q)\Theta(1-G(q)) +\Theta(G(q)-1)$ to be precise where $\Theta(x)$ denotes the Heaviside step function. In our case, its onset value
  $q_{\rm merge}$ 
 may be estimated  to be $q_{\rm merge}\simeq 0.42$ by requiring $G(q_{\rm merge})=1$. This value lies then in the transition region between $q\gg 1$ and $q \ll 1$.
Thus, in the classical regime of $q\gg 1$, their merging dynamics 
will be a main concern, which requires separate studies  \cite{Guzman:2018evm, Guzman:2019bsz}.

Another relevant issue in our collisions of solitons is regarding the effect
of so-called dynamical friction \cite{Chandrasekhar:1943ys}. To study dynamical  friction with our fuzzy DM,   one considers a mass $M_{\rm src}$ located at $r=0$, upon which a Schr\"odinger wave  is incident with a wave number $k= 1/\lambdabar(v)$ and a DM density $\rho$. This then corresponds to a Coulomb-type scattering problem, whose details are analyzed in \cite{Hui:2016ltb,Lancaster:2019mde}. We assume that the mass may be in general distributed over the radius scale $\ell_{\rm size}$ and the point-mass limit then corresponds to $\ell_{\rm size}=0$.
In the frame where the mass is moving through the medium of DM particles,
due to the gravitational interactions between the moving body and the DM particles, the wake behind the moving body is generated and the resulting over density of DM particles exerts a drag force on the body leading to
the effective frictional force
\be
F= -4\pi \rho \left( \frac{G M_{\rm src}}{v}\right)^2 C(q_{src},q_{r},q_{\rm size})
\ee
where one evaluates the drag force at a surface of radius $r$ with
$q_r=\lambdabar(v)/r$, $q_{src}=\lambdabar(v)M_{src}/(f_0 \ell_c{\cal M})$
 and $q_{\rm size}=\lambdabar(v)/\ell_{\rm size}$.
 In application of the present case, the incident wave is replaced by our solition with relative velocity $v$ and the central body may be  either a DM solition or star components distributed over the soliton size scale. Then we naturally take $q_r\sim q_{\rm size}\sim q$ for either for DM-DM or S-DM collisions. Since the frictional force is exerted roughly for $r_{1/2}(M)/v$, the contribution of dynamical friction to the relative change $|\Delta v|/v$
 may be estimated as
 \be
 \frac{
 |\Delta v|}{v}\Big{\vert}_{df}\sim \frac{3 f_0^2 M_{src}}{2M} q^4 C(q_{src},q,q)
 \label{dfratio}
 \ee
 where we take $\rho \sim \frac{1}{2} M/(\frac{4\pi }{3}r^3_{1/2})$ using the definition of the half mass radius. For the pointlike source with $\ell_{\rm size}=0$ and for $q_{src} \ll 1$, the expression $C$ becomes \cite{Hui:2016ltb}
 \be
C(q_{src},q_{r},\infty) \sim {\rm Cin}\left(\frac{2}{q_r}\right)
+{\rm Si}\left(\frac{2}{q_r}\right)+O(q_{src})
\label{cinsi}
 \ee
 where ${\rm Cin}(x)= \int_0^x \frac{1-\cos t}{t}dt$ and ${\rm Si}(x)= \int_0^x\frac{\sin t}{t}dt$.

 Now, for the two soliton collision in the above, we take $q_{src}=q$ and then using (\ref{cinsi}) the factor $C(q,q,\infty)$ in the point-mass limit may be estimated as
 $ 1.33 <C(q,q,\infty) < 2.40$ for our simulation data set. If this were the case, then indeed the leading correction to the relative change (\ref{leadingratio}) would be the dynamical friction contribution in (\ref{dfratio}) instead of the $q^5$ term in (\ref{deltav}). However the fact $\ell_{\rm size}\sim r_{1/2}$ and hence $q_{\rm size}=q$ plays a crucial role
 and in this case $C$ will be reduced at least by a factor of $1/100$ for our simulation data set (See Figure 3 of Ref.~\cite{Lancaster:2019mde} for the estimation of finite-size effect). Hence in this case, the dynamical friction contribution to the relative change will be negligible compared to the $q^5$ correction term.

 To check the above claim numerically, we first try the fit of our simulation data
 with $G'=A q^\lambda(1+ B_{d}\, q)$. But the fit does not work due to
 the optimization failure. We then try the fit with $G'=A q^3(1+ B_{d}\, q+ B q^2)$ and find $A=5.53\pm 0.37$, $B_{d}=-2.01\pm 0.39$ and
 $B=13.60\pm0.30$. Hence indeed the magnitude of $B_{d}$ is relatively small but more importantly its signature is negative, which is not allowed by definition. These results seem to indicate that the dynamical frictional contribution to the relative change in velocity is indeed negligible.

Let us now consider the dynamical frictional contribution to the relative change in a collision of a soliton (DM) and its star components (S), which will be useful in understanding the offset ratio below. In this case, the dynamical friction gives us the leading contribution to the relative change.
For its estimation, we use star components of typical dwarf spheroidal galaxy (soliton in our case), whose total mass is typically about $1\%$ of the soliton mass. Thus we take
$M_{src}/M \sim 0.01$. In this case, $C(q/100,q,q)\sim C(q,q,q)$,\footnote{ $q_{src}\ll 1$ corresponds to the quantum regime in which  $q_{src}$ dependence in $C$ can be ignored.  See Figure 2 of Ref.~\cite{Hui:2016ltb} in this regard.} but one has an extra suppression factor $M_{src}/M \sim 1/100$ in the relative change in (\ref{dfratio}). Hence we conclude that the relative change of S-DM collision is negligible compared to that of DM-DM collision.

As we mentioned already, in this note, we are mainly concerned with the offset (difference) between  S-DM and  DM-DM
collisions in galaxy cluster dynamics. Precise determination of this offset in a generic situation is  a rather complicated problem. In case of head-on collision assuming two colliding masses are equal to each other, the above estimation of the $\Delta v$ will give a good approximation of the offset especially in the large velocity limit. There may be  some complications in general. One is, for instance,  an off-center collision problem with a finite impact parameter. The masses of two bodies may be unequal and so on. However, if one is mainly concerned with the offset part only, the above  may serve as a good effective description since our purpose in the next section is a rough estimation of the offset in colliding galaxy clusters. Of course to solve the problem precisely, a full-fledged numerical simulation of multi  galaxy cluster dynamics is needed, which is beyond scope of the current studies. In conclusion, we shall use the above
result (\ref{deltav}) as an effective measure of the offset part of the collision in the next section.

\section{Collisions of galaxy clusters}\label{sec3}
Typical galaxy clusters have about $50\sim 1000$ galaxies,  hot gas
 between the galaxies, and DM halos.
 In typical CDM models DM particles are presumed to be collision-less.
 Stars  can be treated as effectively
 collision-less particles too because the stars are sparse as mentioned previously.
Thus, we expect stars and DM to move together under gravitational influence  during cluster collisions,
 while gases interact with each other and lag behind.
  This expectation was confirmed for the collisions with high collision velocities as in
  the bullet cluster (1E0657-56)~\cite{bullet}.
However, in collisions with low collision velocities as in Abell 520~\cite{abell520}
there are offsets between the locations of galaxies (stars) and the peaks of DM density.
In the self-interacting DM models this discrepancy can be resolved if DM has a velocity dependent cross section.

Using the results of above section we here suggest that in the fuzzy DM model
the galactic DM halos can have non-negligible  offsets from the stars
even without self-interaction ~\cite{Paredes:2015wga,Guzman:2016peo}.
This is due to the fact that stars and fuzzy DM satisfy two different 
equations; Newton's second law and the Schr\"{o}dinger-Poisson
equation.
 We  consider the collision of two equal mass fuzzy DM solitons representing two  galactic DM halos. Since
  we can approximate the collision of two galaxy clusters by  collisions
of many individual galaxies surrounded by the galactic DM halos, the soliton collisions can give us some
insights for the DM-offset problem.

For this purpose, we introduce here an effective cross section weighted by
the relative change of velocity $|\Delta v|/v$
\be
\sigma_{\rm eff}(v)=\frac{|\Delta v|} {v} 4\pi r^2_{\frac{1}{2}}=G(q)\, 4\pi r^2_{\frac{1}{2}}
\ee
where we take $G(q)=A q^3$ as a leading approximation. This effective cross section is viewed as an extra contribution above the pure gravitational one;
the effective  change in velocity is mainly due to the quantum gravitational cooling effect 
as was shown previously.
When a probe soliton in one cluster is passing through the DM soliton distribution of the other cluster, the change in its velocity
is governed by\footnote{Interestingly, with $G(q)=A q^3$, $\sigma_{\rm eff}/M$ is independent of the soliton mass $M$.}
\be
dv=-v \frac{\sigma_{\rm eff}(v)}{M} \rho({x})dx
\ee
where $M$ is the soliton mass and $\rho({x})$ denotes the DM mass density of the other cluster. When the velocity change is relatively small, the above
equation may be integrated
as\footnote{The approximation here breaks down when ${|v-v_I|}/{v}$ becomes ${\cal O}(1)$. In this regime, one has in general further enhancement of the offset ratio.}
\be
\frac{v-v_I}{v} \simeq -\frac{\sigma_{\rm eff}(v)}{M}\int^x_{x_I} dx'\rho({x'})
\ee
where $x_I$ and $v_I$ are respectively denoting the initial relative position and velocity of our probe soliton.  The offset distance may be obtained as
\be
{x_{\rm off}}
= 
\int^x_{x_I} dx'\frac{v(x')-v_I}{v(x')}
\simeq -\frac{\sigma_{\rm eff}(v)}{M}
\int^x_{x_I} dx'\int^{x'}_{x_I} dx''\rho({x''})
\ee

We consider an enhancing factor 
due to local velocity distribution in each cluster defined by
\be
\frac{E(v)}{v^3} \equiv  \int dv'  \frac{1}{{v'}^3}p(v')
\ee
where $p(v)$  denotes the probability distribution function of soliton velocity in each cluster.
This leads  to the definition of enhanced effective cross section
\be
\bar{\sigma}_{\rm eff}(v)=\sigma_{\rm eff}(v) E(v)
\ee
With this enhancement factor included, our final expression for the offset ratio becomes
\be
\frac{x_{\rm off}}{x_F-x_I} \simeq   -\frac{\bar\sigma_{\rm eff}(v)}{M}\frac{1}{x_F-x_I}
\int^{x_F}_{x_I} dx'\int^{x'}_{x_I} dx''\rho({x''})
\ee
where $x_F$ denotes the final position of our probe soliton. For simplicity we assume a constant distribution between $v_\pm=v\pm\delta v$.
Then, one finds
\be
E(v)=\frac{1}{(1-\frac{\delta v^2}{v^2})^2}
\ee
Using our parameter set given in the previous section, the enhanced effective cross section may be evaluated as
\be
\bar\sigma_{\rm eff}/M = 231.6 \, \frac{v_c^3}{v^3 (1-\frac{\delta v^2}{v^2})^2}{\rm cm^2/g}
\ee
which is independent of the soliton mass $M$ as noted before. As an illustration,
with $(v,\delta v)=(500,450),(1000,950),(4500,950)$ all in the unit of ${\rm km/s}$, one finds respectively the values for the enhanced cross section as
$\bar\sigma_{\rm eff}/M = 1.15, \ 0.546, \ 6.24\times 10^{-5} {\rm cm^2/g}$.
This indeed shows very strong velocity dependence of the cross section.

For a further illustration of the offset ratio, we choose the averaged mass of a soliton
$M_{sol}$ as $2\pi f_0 \times 10^7 M_\odot \simeq 2.47\times  10^8 M_\odot$.
For a typical cluster we choose $M_{cl} \simeq 5 \times 10^{14}M_\odot$
and the number of the soliton in the cluster is then $N_{sol}=M_{cl}/M_{sol}\simeq 2.03 \times 10^6$.
We choose 
the typical cluster size $L_{cl}$ as $5\,{\rm Mpc}$ with $x_F-x_I=2L_{cl}$.  For the sake of illustration, we further assume the density profile $\rho(x)$ such that
\be\label{overd}
\frac{1}{x_F-x_I}
\int^{x_F}_{x_I} dx'\int^{x'}_{x_I} dx''\rho({x''}) \simeq 10\, (x_F-x_I)  \bar{\rho}
\ee
where $\bar{\rho}\simeq M_{cl}/L_{cl}^3$ denotes the average mass density of each cluster.
The factor $10$ on the right hand side of (\ref{overd}) is from an appropriate choice of the density profile
in which one has some  highly over-populated regions. Note that typical galaxy cluster has a core radius $r_c\simeq 0.2 {\rm \,Mpc}$ and core density $\rho_c\simeq 10^{15} M_\odot/{\rm Mpc}^3$~\cite{Bahcall:1995tf}. Then for the above three choices of $(v,\delta v)$,
one finds 
the values for the offset ratio as
$|x_{\rm off}|/(x_F-x_I) = 0.0961,\  0.0456, \ 5.21\times 10^{-6}$, respectively.
Using the result one can roughly estimate the spatial offset in galaxy clusters.
Since stars can be treated as point particles satisfying Newton's equation during cluster collisions, we expect $x_{\rm off}$
is of order of the offsets between DM and stars.
 Assuming $(x_F-x_I)\simeq r_c \simeq 0.2 {\rm \,Mpc}$ gives
 $|x_{\rm off}|= 0.0192{\rm \,Mpc},\ 0.00912{\rm \,Mpc},\  10^{-5}{\rm \,Mpc}$, respectively.
 These values are similar to the typical observed offsets $O(10^{-2}{\rm \,Mpc})$ between peaks of DM and
 stars in galaxy clusters ~\cite{Harvey:2015hha}.


Before our work the offset was shown only by numerical works. Albeit simple,
we present a semi-analytic explanation for the offset using gravitational cooling for the first time.
Our estimation in this section is admittedly very crude, and only  order of magnitude results are
worthwhile to note at this stage.

\section{Conclusions}\label{sec4}

Before this study the quantum effects in fuzzy DM collisions have been studied mainly by numerical methods.
In this semi-analytic study we have shown that solitons of the fuzzy DM have a strong velocity dependent effective cross section
mainly from quantum  gravitational cooling,
 which is inversely proportional to the collision velocity cubed.
 It might explain the offsets between stars (member galaxies) and dark matter in collisions of various galaxy clusters. It might also explain  ultra diffuse galaxies which have an extremely low luminosity \cite{udgalaxy}.
 Our semi-analytic analysis is new, but it is an order of magnitude estimation for the simplest head-on collision problem.
 More general cases with less symmetries require further numerical and analytical studies.
Since the Schr\"odinger-Poisson system is used in various research fields from fundamental quantum mechanics to condensed matter physics,
our results may be relevant to these fields too.

\subsection*{Acknowledgement}
 D Bak was supported in part by the 2021 Research Fund of the University of Seoul.
JW Lee was  supported in part by NRF-2020R1F1A1061160.   S Park was
supported in part by
NRF Grant 2020R1A2B5B01001473, by  Basic Science Research Program
through National Research Foundation funded by the Ministry of Education
(2018R1A6A1A06024977).

\end{document}